\documentclass[aip,apl,amsfonts,amsmath,amssymb, floatfix]{revtex4-1}
\usepackage[utf8]{inputenc}
\usepackage{textcomp}
\usepackage{appendix}
\usepackage{graphicx}
\usepackage{color}

\begin{document}

\title{Clean, Robust Alkali Sources by Intercalation within Highly-Oriented 
Pyrolytic Graphite}
\author{Rudolph N. Kohn, Jr.}
\email{rudy.kohn@sdl.usu.edu}
\affiliation{Space Dynamics Laboratory, Albuquerque, New Mexico 87106, USA}
\author{Matthew S. Bigelow}
\affiliation{Applied Technology Associates, Albuquerque, New Mexico 87123, USA}
\author{Mary Spanjers}
\affiliation{Air Force Research Laboratory, Kirtland AFB, New Mexico 87117, USA}
\author{Benjamin K. Stuhl}
\affiliation{Space Dynamics Laboratory, Albuquerque, New Mexico 87106, USA}
\author{Brian L. Kasch}
\author{Spencer E. Olson}
\affiliation{Air Force Research Laboratory, Kirtland AFB, New Mexico 87117, USA}
\author{Eric A. Imhof}
\affiliation{Space Dynamics Laboratory, Albuquerque, New Mexico 87106, USA}
\author{David A. Hostutler}
\author{Matthew B. Squires}
\affiliation{Air Force Research Laboratory, Kirtland AFB, New Mexico 87117, USA}

\begin{abstract}

We report the fabrication, characterization, and use of rubidium vapor 
dispensers based on highly-oriented pyrolytic graphite (HOPG) intercalated with 
metallic rubidium. Compared to commercial chromate salt dispensers, these 
intercalated HOPG (IHOPG) dispensers hold an order of magnitude more rubidium in 
a similar volume, require less than one-fourth the heating power, and emit less 
than one-half as many impurities. Appropriate processing permits exposure of the 
IHOPG to atmosphere for over ninety minutes without any adverse effects. 
Intercalation of cesium, potassium, and lithium into HOPG have also been demonstrated in 
the literature, which suggests that IHOPG dispensers may also be be made for 
those metals.

\end{abstract}

\maketitle

\section{Introduction} \label{sec:intro}

Alkali metals serve as the atomic backbone for a wide variety of physics 
experiments. Alkalis' simple electronic structure and strong transitions 
facilitate laser cooling and make them very attractive candidates for atomic 
sensors. In experiments using alkali metals, the on-demand production of a 
clean, dilute vapor is often the first crucial step. The stringent requirements 
of modern cold atom experiments mean that significant improvements in this first 
step can positively impact the rest of the experiment.  This work introduces a new 
architecture for producing extremely pure alkali vapors using intercalated 
graphite, and compares this new architecture to commercially available 
dispensers.

Many methods exist to produce pure, dilute atomic vapors. However, experimental 
parameters often limit which methods are viable. For example, an atomic beam 
experiment might need a source with high flux and directionality, but stationary 
or slow-moving cold atom experiments often depend upon extremely pure, dilute 
vapors, so that the atoms can be trapped and cooled with minimal interference 
from background gas. The desired background pressures are often orders of 
magnitude below the room temperature vapor pressures of the metals, impeding the 
use of pure metallic sources. Other characteristics, such as capacity, 
activation temperature, ease of handling, and total vapor produced frequently 
place other restrictions on the source. 

For the production of pure, dilute alkali vapors, there are two general 
architectures in common use. The first uses ovens, usually emitting effusive 
beams, which offer high purity and capacity (grams) and low activation 
temperature ($\sim$ 100 \textcelsius\ for rubidium). However, simple ovens 
produce enough vapor to have deleterious effects on pumps and vacuum quality. 
\cite{Lin2009} Negative effects on vacuum can be mitigated, but generally at the 
expense of increased complexity. The large quantities of 
pure alkali used can present difficulties for safe handling and disposal, as 
well. \cite{Walkiewicz2000}

The second general architecture uses chromate salts and non-evaporable getter 
(NEG) material to produce a reasonably pure alkali vapor. 
\cite{SAESGetters2007}\textsuperscript{,} \footnote{Various commercial 
equipment, instruments, and materials, as well as their suppliers are identified 
in this paper to foster understanding and reproducibility.  These 
identifications do not imply endorsement by the Air Force Research Laboratory, 
or that other materials are unsuited to these purposes.}  Commercial chromate 
salt dispensers are compact and easy to handle, but they have relatively low 
capacity ($\sim$ 10 mg for a 30 mm long dispenser), much higher activation 
temperatures ($\sim$ 500 \textcelsius\ for rubidium chromate), and can emit 
significant quantities of unwanted gas under certain conditions. A typical 
chromate salt dispenser contains NEG material to inhibit the release of unwanted 
gases. However, prolonged periods at room temperature can collect unwanted gases 
in the NEG material or on the nearby chamber walls, and improperly degassing 
before activation can contaminate the source. \cite{Fortagh1998}    We have 
observed that extended periods at room temperature produce a measurable pressure 
spike when the chromate dispenser is next heated, which can be traced to either 
gas adsorbed onto the steel container of the dispenser or absorbed into its NEG 
material, or adsorbed onto chamber surfaces near the dispenser, which are heated 
during activation. 

We present an alternative source of alkali vapor which compares favorably to the 
chromate salt architecture in size, ease of handling, and ease of activation 
while improving on the purity of their output and increasing their capacity to 
$\sim$ 100 mg in a similar volume. Highly-oriented pyrolytic graphite (HOPG), 
essentially graphite with a high degree of internal order, can absorb relatively 
large amounts of foreign chemicals between its graphene-like layers. This 
behavior, known as intercalation, has been studied since at least the 1940s. 
\cite{ZAAC, CROFT1951}  Graphite and HOPG have frequently been used as getters 
for alkali vapors because of this behavior. \cite{Jefferts2002, Levi2003}  The 
characteristics of intercalated HOPG (IHOPG) as a dispenser have been examined 
in the context of vapor deposition, \cite{Baumann1985} but, to the best of our 
knowledge, no examination of its compatibility with ultra-high vacuum (UHV) or 
modern cold-atom experiments has been made. Many materials are known to 
intercalate into HOPG, but of special interest here are the four alkalis known 
to intercalate relatively easily:  lithium, potassium, rubidium, and cesium. 
\cite{Guerard1975, Shu1993, Jungblut1989, Salzano1965, Salzano1966, Aronson1968} 
 Studies have shown that sodium is much more difficult to intercalate into 
graphite, \cite{Asher1959} therefore it seems unlikely that reliable dispensers 
can be made for sodium. Although examples of cesium, potassium, and lithium 
intercalation are present in the literature, the work described here uses 
rubidium exclusively. We devised a method to reliably intercalate HOPG with 
rubidium, make the IHOPG relatively stable in atmosphere, and controllably 
dispense rubidium vapor under vacuum.

We describe the methods for producing rubidium IHOPGs in Section \ref{sec:fab}. 
The apparatus used to compare the dispensers is detailed in Section 
\ref{sec:apparatus}, and then, two comparisons are made. The first, in Section 
\ref{sec:steady}, compares the purity of the output vapor of both dispensers in 
a steady-state configuration. The second, in Section \ref{sec:adsorb}, compares 
the undesired gas which accumulates on, in, or near the dispensers after 
extended periods at room temperature. In both cases, the IHOPG dispenser 
produced less undesired gas than the chromate dispenser.  In general, heating 
dispensers emits waste gases, either from the dispenser itself or from gases 
adsorbed onto nearby walls.  Emitted rubidium interacts with background gas, 
causing a getter effect.  Over the range of rubidium emission rates we tested, 
the waste gases produced by heating the chromate dispenser consistently 
overwhelmed this getter effect.  However, at sufficiently high rubidium output 
rates the waste gases produced by heating the IHOPG were so inconsequential that 
the pressure in the chamber actually decreased. Aside from the IHOPG dispenser 
used for these tests, another IHOPG dispenser has been successfully integrated 
into a cold atom experiment loading grating magneto-optical traps (MOTs). 
\cite{Imhof2017}  Another system in current daily use contains an IHOPG for 
dispensing rubidium, and regularly produces Bose-Einstein condensates.

\section{Fabrication and Operation} \label{sec:fab}

Alkalis are intercalated into HOPG by placing a heated sample of HOPG in close 
proximity to vapor at high enough ($\gtrsim$mTorr) pressure. The heat allows 
alkali atoms to diffuse between its graphene-like layers. Specific temperatures 
are given for rubidium. Cesium has a very similar melting point and vapor 
pressure, so the temperatures required may also be very similar. Potassium and 
lithium, however, melt at much higher temperatures and have a much lower vapor 
pressure, so higher temperatures will almost certainly be required.

The procedure detailed below reliably produces rubidium IHOPG dispensers with 
about 1 mg of rubidium per mm$^3$ of dispenser. The initial samples of HOPG were 
7 x 7 x 1 mm, but successful intercalation increased their volume by a factor 
between 2 and 3. Several optional steps, detailed in Subsection 
\ref{ssec:handle}, improve ease of handling. 

The structure of the original HOPG affected how reliably they could be 
intercalated. HOPG is typically graded by crystallographic order,  
characterized by two parameters:  mosaic angle and grain size. Mosaic angle is 
a measure of the dispersion of the angles of crystallites in the sample. 
\cite{Ohler:vi0133,Bremer1993,Stout1968}  The grain size is typically measured 
in microns or millimeters and describes how far apart, on average, grain 
boundaries can be found. In the early stages of our work, we experimented with 
samples with different levels of order and found that samples with higher 
crystallographic order loaded and dispensed more reliably. The dispensers 
described here were all produced from HOPG samples with mosaic angles of 0.8 
$\pm$ 0.2 degrees, and with average grain sizes between 0.5 and 1.0 millimeters. 

We prebake all of our HOPGs at 250 \textcelsius\ under high vacuum to eliminate 
surface impurities and degas the sample. 48-72 hours at this temperature, with a 
turbomolecular pump to maintain vacuum, is sufficient to bake out a reasonably 
clean HOPG sample, removed from its packing material and handled with gloves. 
After the prebake, we remove the sample to a dry nitrogen atmosphere and add 
elemental rubidium into the chamber.  The glove box we use for this purpose is 
fed dry nitrogen from a dewar source, but any inert gas (e.g. argon) would 
likely work just as well.  In practice, we usually pour a few drops of molten 
rubidium onto the HOPG, but other delivery methods, such as small pieces of 
solid rubidium dropped into the chamber should work just as well, since the 
intercalation process is driven by heating the HOPG and exposing it to vapor, 
\cite{SalzanoInst1965, Nixon1968} which will be present when the chamber is 
heated in either case. Before heating, we attach the chamber to an oil-free 
roughing pump, reducing the background pressure to the milliTorr range. The 
rough vacuum removes unwanted gases in the chamber, which can impair 
intercalation. We then seal off the chamber containing the HOPG and rubidium 
under rough vacuum and heat the chamber to 125-150 \textcelsius\ for at least 48 
hours to intercalate. 

After 48 hours, we turn off the heaters and remove the cooled chamber to a dry 
nitrogen atmosphere to examine the HOPG. Successfully intercalated HOPGs 
dramatically increase in thickness, with a 1 mm thick HOPG swelling to 2 or 3 mm 
after intercalation. This expansion is strong evidence that the process creates 
high stresses in the HOPG sample. We have observed samples that broke into 
several pieces during intercalation, and an attempt to load rubidium into 
graphene foam reduced the sample to dust. Based on these observations, we infer 
that larger grain size inhibits structural damage to the HOPG.

The structure and dimensions of IHOPG are fairly well known \cite{Salzano1966}. 
IHOPG has several different stable structures with different stoichiometric 
ratios. The structure with the most intercalated material has the formula 
XC$_8$, corresponding to 0.89 g of rubidium for each gram of carbon. However, 
our loaded samples usually gain more mass than this, with a 110 mg HOPG gaining 
100-220 mg of rubidium. The expansion in size is also larger than predicted by 
the expected structure:  the known thickness difference between pure and 
maximally intercalated HOPG is 68\%--much less than the 100-200\% increases 
observed in our samples. In addition, IHOPGs exposed to air for long periods 
tend to expand further, with layers splitting apart as the rubidium oxidizes.  
These observations strongly suggest that additional rubidium is making its way 
between the layers of the HOPG, pushing them further apart and adding more 
rubidium to the dispensers than expected in a pure intercalation.

Once the HOPGs are loaded, they are placed in a vacuum chamber and heated in 
order to liberate the intercalated rubidium. The IHOPGs begin emitting rubidium 
vapor when heated over some activation temperature, which varies somewhat from 
sample to sample. Typical activation temperatures range from 125-160 
\textcelsius, and seems to roughly correlate with the amount of intercalated 
rubidium.  Additional heating over the activation temperature increases the 
emission rate. Dispensers rapidly plate rubidium onto nearby glass at 
temperatures of 250 \textcelsius. A newly loaded dispenser held at 250 
\textcelsius\ depleted itself in about 72 hours, implying a rate of 2-3 
milligrams per hour at that temperature. A different IHOPG sample emitted no 
measurable rubidium over a 12 hour period at 150 \textcelsius, but at 170 
\textcelsius, it produced enough rubidium to observe laser-induced fluorescence 
in a small chamber after about 10 minutes, suggesting an output rate of about 
0.5 nanograms per hour, taking into account the laser power and the sensitivity 
of the infrared scope. Our attempts to measure rubidium vapor emitted from 
IHOPGs at room temperature under vacuum have all been below the detection limit.

We attempted to heat the IHOPGs above the activation temperature with several 
methods. Of these methods, we had success with two:  conductive heating from 
outside the chamber and inductive heating across a glass wall. For conductive 
heating, a resistive tape outside the chamber heats the IHOPG through the 
chamber wall. This method allows easy monitoring of the IHOPG temperature with a 
thermocouple, which is useful for characterizing a sample's activation 
temperature. Conductive heating might also be accomplished by attaching the 
IHOPG to a heating element inside the chamber. For inductive heating, a wire 
coil outside the chamber, oriented in the same plane as the layers of the IHOPG 
and carrying a large, rapidly oscillating current, heats the sample by inducing 
eddy currents in the graphite. In a glass chamber, the IHOPG can be epoxied to a 
chamber wall or simply rest on the chamber bottom. The distance between the coil 
and the IHOPG controls the temperature and emission rate, though measuring the 
absolute temperature using this method is difficult.

\subsection{Improving Ease of Handling} \label{ssec:handle}

Although the dispenser can be used immediately after loading, it usually leaves 
the intercalation chamber coated in a layer of metallic rubidium. Failing to 
remove the surface rubidium results in rapid reactions with air and moisture 
upon exposure, and can cause structural damage to the IHOPG after only a few 
minutes. A few optional steps greatly ease handling in atmosphere. First, we 
transfer the IHOPG into a clean glass vacuum chamber and heat it with a heater 
tape to between 100 and 120 \textcelsius, with a turbomolecular pump maintaining 
vacuum. We maintain the temperature below the activation temperature to remove 
surface rubidium without dispensing from between the layers. Depending on the 
amount of surface rubidium, this takes 24-48 hours. We gauge the depletion of 
surface rubidium by observing laser fluorescence on the D2 line at 780 nm.  
Heated surface rubidium produces enough vapor to observe laser-induced 
fluorescence when a near-resonant laser is passed through the chamber.  When 
this fluorescence disappears, the surface rubidium has been depleted and it is 
safe to move on to the next step.

After removing the surface rubidium, we slowly raise the temperature of the 
IHOPG to find the activation temperature. A thermocouple provides temperature 
data, and a resonant laser beam passing through the tube produces fluorescence 
when the IHOPG begins to emit rubidium.  Once we observe laser-induced 
fluorescence, we note the current temperature as the activation temperature and 
raise the temperature an additional 40 \textcelsius\ for 8 hours. The goal of 
this 8 hour period is to  deplete the rubidium at the edges of the IHOPG, which 
we suspect reduces its availability to react with local atmosphere. Samples 
treated in this way have been handled in air for 90 minutes or more with no 
visible changes. Without this second step, the sample turns grey as a coating of 
rubidium hydroxide forms on the surface. Samples with this coating have still 
been used to load a MOT. \cite{Imhof2017} However, these reactions are usually 
best avoided, as they may cause physical damage, making heating the IHOPG more 
difficult. For example, a crack could impair inductive heating, or disconnect 
the dispenser from its heater.

\section{Apparatus} \label{sec:apparatus}

\begin{figure}
 \includegraphics[width=8.5cm]{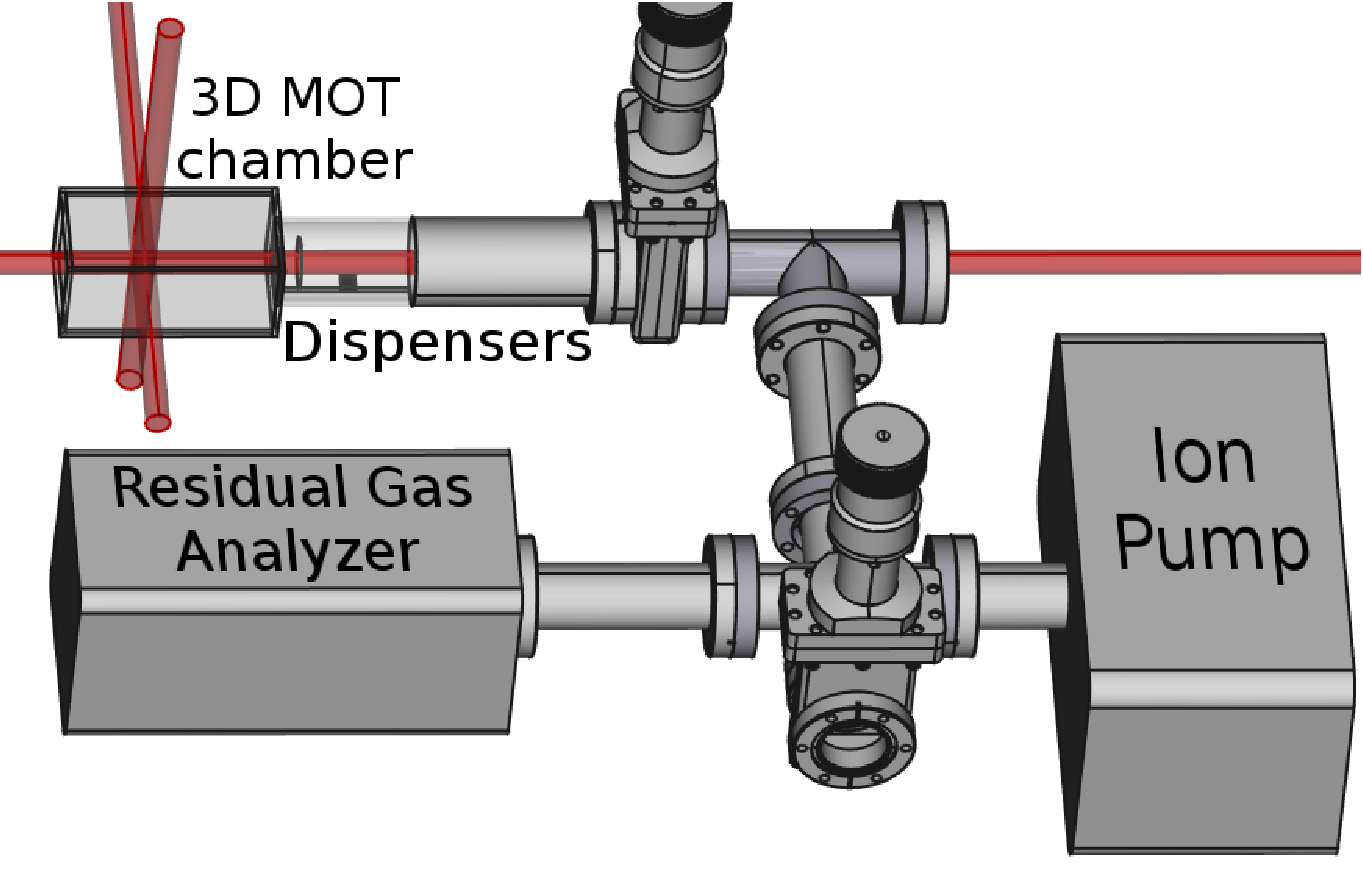}
 \caption{The chamber used for the experiments in Sections \ref{sec:steady} and 
 \ref{sec:adsorb}. The two different dispensers rest near each other in the 
 glass neck at the top of the figure. The rubidium fluorescence is measured 
 at the crossing of the three beams, pictured in red. The retro-reflecting 
 optics for the 3D MOT, the magnetic quadrupole coils, and the inductive 
 heating coils are not shown.}
 \label{fig:chamber}
\end{figure}

In sections \ref{sec:steady} and \ref{sec:adsorb}, we will discuss two 
comparisons between IHOPG dispensers and commercial rubidium chromate 
dispensers. This section describes the apparatus used to perform both of those 
comparisons. A schematic view of the chamber is pictured in Figure 
\ref{fig:chamber}. On the left, a rectangular glass chamber provides optical 
access for a 3D MOT (red laser beams). The IHOPG and a chromate dispenser were 
located near each other, in the cylindrical glass neck between the glass and 
steel parts. A sputter-ion pump and a residual gas analyzer (RGA) were attached 
to the steel part of the test chamber. The RGA measured partial pressures using 
a quadrupole mass analyzer, but measured the total pressure using a separate ion 
gauge filament. The unit used in these experiments was new, and for a measured 
total pressure of $1.3 \times 10^{-8}$ Torr the sum of the pressure peaks from 
the mass spectrometer was $9 \times 10^{-9}$ Torr. The RGA measured mass/charge 
ratios out to 90 AMU/e.

The two types of dispensers were oriented in perpendicular planes to permit 
selective inductive heating. The power transferred by an inductive heater is 
highly dependent on the spatial orientation of the coils and the object to be 
heated.  The inductive heater coil for the chromate dispensers was wrapped 
around the cylindrical glass chamber and moved back and forth along the 
cylindrical section, changing its distance from the chromate dispenser to 
control its temperature. A different coil was placed under the IHOPG and moved 
up and down to control its temperature. The linear translation stages used to 
move the coils had 8 mm of travel. Due to space constraints, only one coil was 
in position at any given time.  The two dispensers were only a few millimeters 
apart in the chamber, but based on the effective ranges of the inductive heaters 
(less than 8 mm), and on how strongly the orientation of the heaters affects the 
level of heating, we are confident that the heating of each individual dispenser 
did not significantly heat the other.  The orientation and size of the coil for 
the chromate dispenser makes the induced fields over the volume of the IHOPG 
very small.  Similarly, the coil for heating the IHOPG was several millimeters 
from the chromate dispenser and was much smaller than the chromate dispenser, 
meaning that it was extremely unlikely to induce significant currents in the 
chromate dispenser.  In addition, the different activation temperatures make it 
highly unlikely that the chromate dispensers would be anywhere near activation 
when the IHOPG was heated. When the chromate dispensers were heated, the heating 
of the glass in the area of the IHOPG was minimal.  The lack of local heating is 
attributed to the use of Litz wire for the chromate heater coil, which helped 
reduce the resistivity of the coil at the high frequencies produced by the 
inductive heater.

The rectangular glass chamber admitted three perpendicular, circularly-polarized 
laser beams. These three primary beams all contained light 13 MHz 
red-detuned from the 5 $^2$S$_{1/2}$, F = 2 to 5 $^2$P$_{3/2}$, F = 3 line in 
$^{87}$Rb. A repumping beam, tuned to the transition between the 5 
$^2$S$_{1/2}$, F = 1 and 5 $^2$P$_{3/2}$, F = 2 resonance, was aligned with the 
beam going from right to left in Figure \ref{fig:chamber}, and could be turned 
on and off independently of the primary beams. The three primary beams had a 
total power of approximately 13 mW, and each beam had a waist of about 7 mm. The 
repump beam had 1.8 mW of light in a similarly sized beam. Each beam was 
retro-reflected through a quarter wave plate to produce the light fields 
necessary for a 3D MOT. Magnetic coils outside the vacuum chamber produced the 
necessary quadrupole field at the intersection of the three beams. The MOT 
ensured that the laser was locked to the same frequency for each run, and gave 
an early indicator of rubidium vapor output.

\begin{figure*}
 \includegraphics[width=16.4cm]{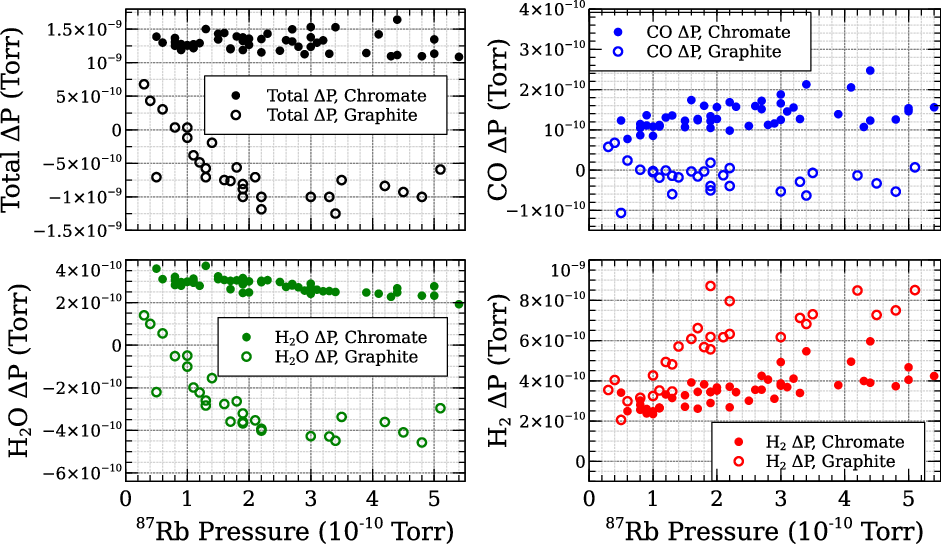}
 \caption{Changes in total pressure and relevant partial pressures at 
 steady-state with each dispenser type. 
 The rubidium vapor density is approximated from fluorescence measurments.  
 The vapor density is in turn determined by dispenser temperature and serves
 as an indicator of the emission rate.}
 \label{fig:deltaP}
\end{figure*}

The sputter-ion pump maintained the chamber pressure near $1.3 \times 10^{-8}$ 
Torr. The chamber was baked out for 5 days at 125 \textcelsius\ using a 
turbomolecular pump to minimize impurities, but the Viton gasket seal on the 
exit valve limited the overall quality of the vacuum.  The RGA measured a 
partial pressure of water vapor of $1.3 \times 10^{-9}$ Torr, with additional 
peaks of hydrogen ($2.7 \times 10^{-9}$ Torr) and nitrogen and carbon 
monoxide ($1.1 \times 10^{-9}$ Torr, at the same mass value). Other, smaller peaks 
were also present at several common mass numbers (e.g. carbon dioxide), but they 
did not significantly change upon heating of either dispenser. Neither 
dispenser, when heated, introduced new significant peaks to the RGA traces. 
Rubidium vapor was strongly attenuated by the chamber walls, never reaching the 
RGA in measurable concentrations.  However, the dispensers were run for periods
of several hours per day for several days before taking data, in order to
align the laser beams and test and optimize the fluorescence measurements.

Before taking data, each dispenser was individually degassed over the course of 
several hours. Each dispenser was slowly heated while the RGA monitored the 
released gases. The MOT beams provided feedback to determine when rubidium 
output began. As the temperature was increased, occasional bursts of output gas 
were measured on the RGA and were allowed to dissipate before further increasing 
the temperature. This process was continued until rubidium output was observed. 
The dispenser being used was then cooled to room temperature, and the background 
pressure in the chamber was allowed to stabilize, as measured on the RGA, before 
starting a run. 

\section{Steady-state Output Comparison} \label{sec:steady}

\begin{figure}
 \includegraphics[width=8.5cm]{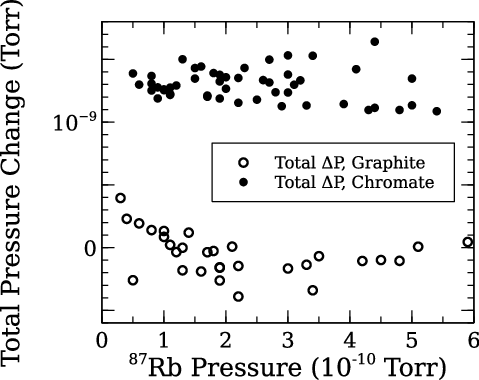}
 \caption{Total pressure comparison, with the graphite dispenser data corrected 
for reactions between water vapor and rubidium. The chromate dispenser data are 
unchanged from Fig. \ref{fig:deltaP}, and are shown for reference only. 
The pressure loss due to dispensed rubidium 
reacting with background water vapor has been added back to the IHOPG 
data, 
using best estimates for the sensitivity differences between the mass 
spectrometer and ion filament, as well as the measured ratio of the 17 AMU peak 
to further refine the correction. These data provide approximate values for the 
expected pressure differences that would be observed in a chamber without 
background water vapor. Of note is that even with the water vapor losses
added back to the IHOPG data, the total pressure increase is near zero, and 
still much less than the observed pressure increases from the chromate 
dispenser, even without correcting those data for the water vapor reaction.}
 \label{fig:wwater}
\end{figure}

Each dispenser was heated individually to a steady-state to measure the output 
gases. With the small amounts of rubidium used, the steel chamber walls were 
never saturated with rubidium, so the rubidium vapor was not detected by the 
RGA. Instead, rubidium density was measured by observing the fluorescence of the 
atoms illuminated by the MOT cooling beams, imaged onto a photodiode. The 
measurements used in Figures \ref{fig:deltaP} and \ref{fig:wwater} were taken 
with the quadrupole field and repump light turned off, as day-to-day variations 
in MOT shape resulted in significant variation in measured MOT fluorescence. In 
contrast, the light from the untrapped atoms was much more repeatable from day 
to day, and the imaging system had a much clearer global maximum during 
alignment.  Between the cylindrical neck where the dispensers were and the 
chamber where the fluorescence was measured, there was a glass plate with an 
aperture.  Neither dispenser had direct line-of-sight access to the fluorescence 
chamber.  Rather, each dispenser produced a local cloud of rubidium, and a 
fraction of the atoms were diffusely reflected by the walls into the 
fluorescence chamber. Since the steel half of the chamber effectively functioned 
as a pump for rubidium, the local rubidium density near the dispensers should be 
roughly proportional to the dispenser's emission rate.  The mean free path in 
the system is expected to be much larger than the size of the system, so the 
atoms that are directed at the crossed laser beams should reach them relatively 
unimpeded.  

Calculations assuming a room-temperature 
distribution of rubidium and using the measured powers and detunings of the 
beams provided a rough estimate of local rubidium pressure as a function of 
observed fluorescence.  For each of the six laser beams, we calculated the
total fluorescence $F$ in photons per second:
\begin{equation}
 F = n \int_{-\infty}^{\infty} \text{d}^3\vec{v} \int_{-0.0036}^{0.0036}
 \text{d}^3\vec{x} \; f(\vec{v}) \; \Gamma(\vec{x},\vec{v}),
\end{equation}
where $\vec{x}$ and $\vec{v}$ are position and velocity, respectively, $n$ is
the $^{87}$Rb density,
$f(\vec{v})$ is the 3D Maxwell-Boltzmann distribution
\begin{equation}
 f(\vec{v}) = \left( 
  \frac{m}{2 \pi k_B T}
 \right)^{3/2}
 \exp(-m |v|^2/2 k_B T),
\end{equation}
and $\Gamma(\vec{x},\vec{v})$ is the rate of spontaneous emission
\begin{equation}
 \Gamma(\vec{x},\vec{v}) = \frac{\gamma}{2} 
 \frac{I(\vec{x})/I_{\text{sat}}}{1 + I(\vec{x})/I_{\text{sat}}+(2(\delta-\vec{k} \cdot \vec{v})/\gamma)^2}.
\end{equation}
$I(\vec{x})$ is the intensity profile of the beam in question, assumed to be a
Gaussian in two dimensions and constant in the third direction, as the beams were
collimated.  $\gamma$ is the excited state linewidth ($2 \pi \times 6.065 \times 10^{6}s^{-1}$),
$\delta$ is the detuning ($-2 \pi \times 13 \times 10^6 s^{-1}$), $\vec{k}$ is the 
beam's wave vector, $m$ is the mass of $^{87}$Rb, $T$ is 300 K, $I_{\text{sat}}$ 
is the saturation intensity 
(1.67 mW/cm$^2$), and $k_B$ is Boltzmann's constant.  The limits of the integral 
for position are determined by the field of view for the detector and its 
optics.  The total number of photons was corrected for solid angle, 
efficiency of the detector, and amplification factor of the detector electronics.
These calculations led to a rough estimate of the fluorescence emitted versus the
number density (pressure) of the rubidium, corresponding 
to about $1 \times 10^{-10}$ Torr of 
$^{87}$Rb per millivolt of fluorescence.  Since the six beams technically formed a three-dimensional 
molasses configuration, similar calculations estimated the additional 
contribution from slow atoms, assuming that all atoms with a velocity lower than 
15 m/s were slowed to the recoil velocity, thereby increasing their time in the 
beams and emitted fluorescence. It should be noted that no effort was made to 
zero the magnetic fields, which will weaken the effect of the molasses.  
However, this overestimate leads to additional fluorescence from the slowed 
atoms that was less than that from the rest of the thermal distribution by about 
three orders of magnitude. The effect of the molasses is therefore negligible.

\begin{figure}
 \includegraphics[width=8.5cm]{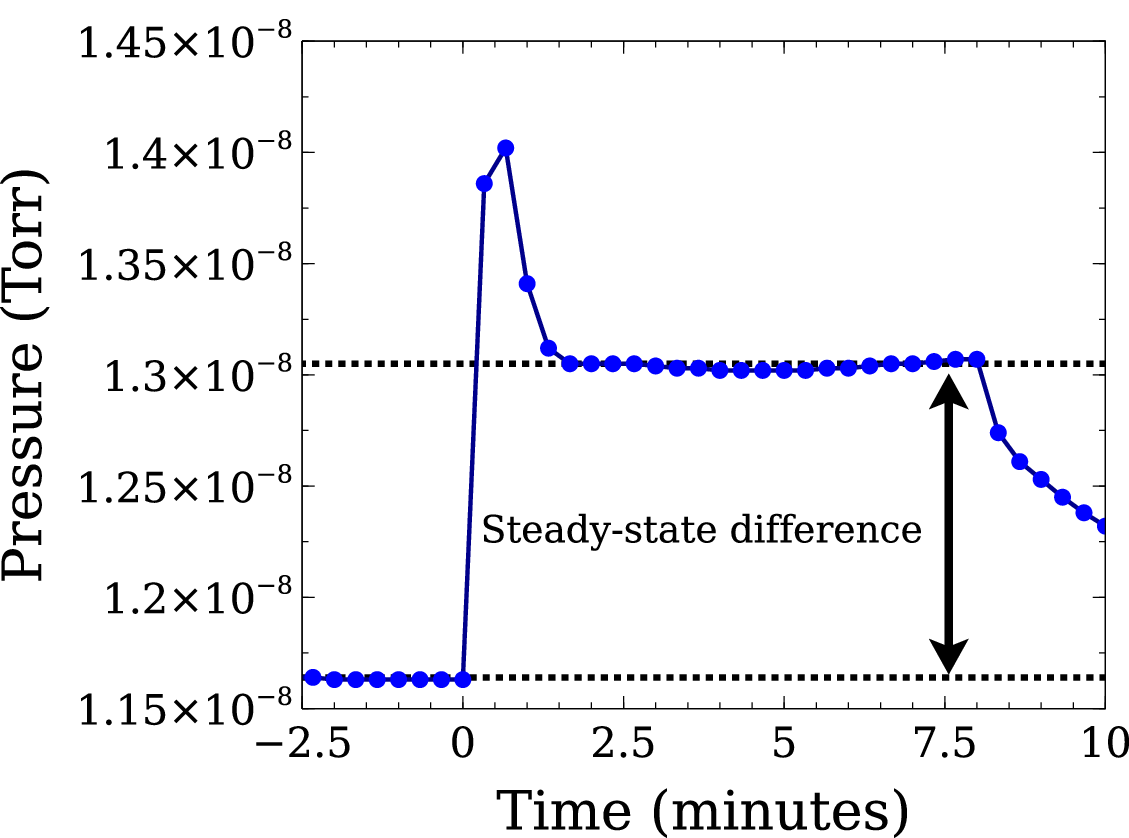}
 \caption{An example of the RGA data used to produce one of the points in Figure 
 \ref{fig:deltaP}.
 These data are from a run of the chromate dispenser, and correspond to total
 pressure measured by the RGA. When the dispenser is heated, the pressure spikes,
 then falls to an equilibrium pressure greater than measured before
 heating. The pressure difference between the measurement before turning
 on the dispenser and at steady state is plotted in Figure \ref{fig:deltaP}
 against the rubidium fluorescence observed at steady-state.}
 \label{fig:ssi}
\end{figure}

Figure \ref{fig:deltaP} details the observed pressure changes as a function of 
rubidium output for the chromate and IHOPG dispensers. Each point represents a 
run in which one of the dispensers was heated until the output gases measured by 
the RGA reached a steady state. In the case of the chromate dispenser, a typical 
run started with a large spike of output gases, followed by a taper over several 
minutes to reach a final plateau (see Figure \ref{fig:ssi}). The chromate 
dispensers typically reached a steady-state after approximately ten minutes, and 
the IHOPG response times were somewhat slower. For low output rates, the IHOPGs 
acted in much the same way, but as the IHOPG temperature and rubidium output 
increased, a qualitative change in the IHOPG behavior occurred due to the 
background water vapor in the chamber.

During runs with the IHOPG at temperatures above a threshold corresponding to 
about 1 mV of rubidium fluorescence, the pressure curves measured by the RGA 
eventually fell below those measured before heating. In these cases, the 
pressure decrease continued very slowly over a fairly long period. The IHOPG 
data shown in Figure \ref{fig:deltaP} were limited to 40 minutes of run time per 
point. The 40 minute duration was chosen to capture the majority of the effect 
while constraining experimental time.  A few tests of the IHOPGs at longer 
times, up to two hours, showed that the 40 minute time limit captured the vast 
majority ($>90\%$) of the effect.

Comparing the total pressure to the water vapor pressure, it is clear that much 
of the pressure decrease is due to the rubidium vapor reacting with background 
water vapor: 
\begin{equation} 
2\mathrm{Rb} + 2\mathrm{H}_2\mathrm{O} \rightarrow 2\mathrm{RbOH} + 
\mathrm{H}_2. 
\end{equation} 
The chamber used in this experiment had a significant background including water 
vapor. However, in a UHV chamber, the getter effect of the rubidium would not be 
present. The well-known characteristics of the RGA measurement make it possible 
to approximate the pressure change in a chamber without background water vapor. 
We take the total pressure data and add the lost water vapor back to each 
individual point. Figure \ref{fig:wwater} shows the data after this correction. 
Instead of simply adding back the lost water vapor, the data are corrected for 
the measured sensitivity differences between the mass filter and ion gauge 
filament, as well as for the measured ratio of the correlated 17 AMU/e peak. 
Correcting for the lost water vapor, the total pressure change is still well 
below that observed from the chromate dispenser over the measured range. As 
noted above, rubidium vapor was attenuated by the chamber walls before reaching 
the RGA. Therefore, the rubidium vapor pressure does not appear in the RGA total 
pressure levels.

The chromate dispensers did not produce a measurable decrease in water vapor
despite the emission of similar amounts of rubidium to the IHOPG dispenser.
We attribute this to the fact that the activation temperature for the chromate
dispenser is much higher, resulting in greater sympathetic heating of nearby chamber
walls and desorption of water vapor from the walls.  Even though
the rubidium from the chromate dispenser was reacting with similar amounts
of water vapor, the amount of water vapor released by heating the chamber
walls exceeded the amount reacting with the rubidium.  A careful examination
of Figure \ref{fig:deltaP} shows that the water vapor output when heating
the chromate dispenser has a weak negative slope as the rubidium output
increased.  The amount of additional desorption as the dispenser temperature
was raised was less than the getter effect of the additional rubidium, but the
high activation temperature meant that there would be significant
heating of the chamber even at very low rubidium output rates.

\begin{figure*}
 \includegraphics[width=16.4cm]{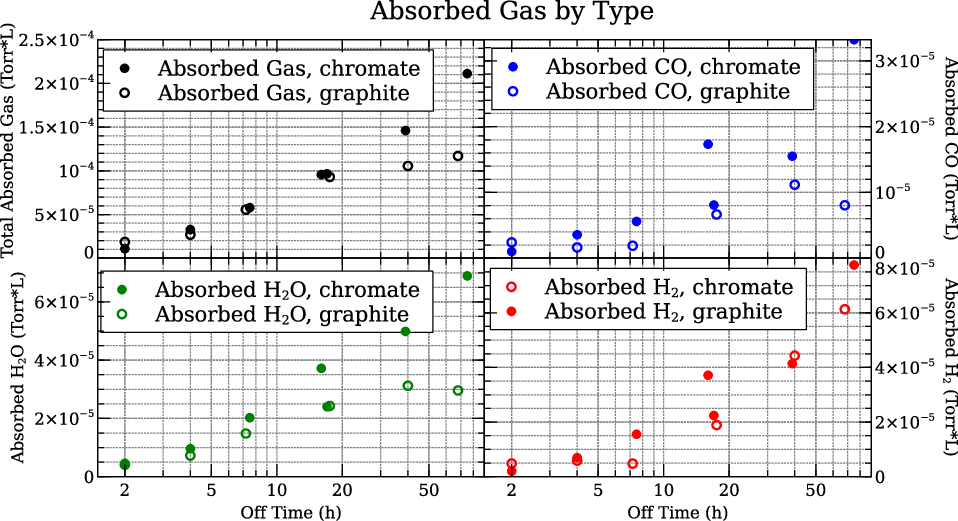}
 \caption{Measurements of absorbed gas in the different dispensers or adsorbed 
 to nearby surfaces heated along with the dispenser. Each point 
is derived by integrating the area under a difference curve calculated by 
subtracting the pressure curve of a one-hour off-time from the chosen, longer 
off-time, and then multiplying by the pumping rate. See Figure \ref{fig:degas}
for a set of example data.}
 \label{fig:abs_gas}
\end{figure*}

In a chamber without background water vapor, the pressure decrease shown in 
Figure \ref{fig:deltaP} would likely not appear, but the corrected data in 
Figure \ref{fig:wwater} suggest that the pressure increase due to undesired gas 
emission from the IHOPG is extremely small. In comparison, the chromate 
dispensers increased total background pressure by about $1.2 \times 10^{-9}$ 
Torr regardless of the level of rubidium output. In the worst case, when the 
IHOPG was producing almost no rubidium, the total pressure increase in the 
steady state was $7 \times 10^{-10}$ Torr. These results show that in the 
configuration used here, the IHOPGs produce, at most, about one-half of the 
waste gases of the chromate dispensers, and compare more favorably at higher 
output rates.  Some of the observed waste gases were almost certainly 
contributed by desorption of gases on nearby walls, but this effect is difficult 
to engineer away, as the heat required for activation, especially for the 
chromate dispensers, is quite large.  In other words, there seems to be a
significant advantage to be gained by using dispensers that require less heat
to activate, as generating less heat mitigates desorption and out-gassing from other
nearby vacuum components.

\section{Gas Absorption Comparison} \label{sec:adsorb}

\begin{figure}
 \includegraphics[width=8.5cm]{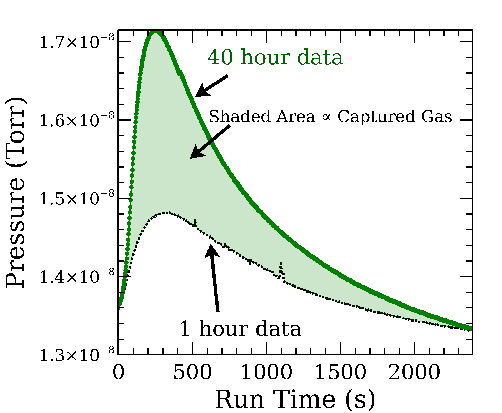}
 \caption{Example of pressure data used to produce the 40 hour point for IHOPG 
gas absorption measurement. The lower curve shows the total pressure as a 
function of time when the IHOPG was heated after 1 hour at room temperature. The 
upper curve shows pressure data collected when the IHOPG was similarly heated after 40 hours at room 
temperature. The area difference between the two curves is used to determine the 
amount of gas absorbed by the dispenser during the 40 hour period. Each point in 
Figure \ref{fig:abs_gas} comes from one such comparison.}
 \label{fig:degas}
\end{figure}

UHV systems have extremely stringent requirements on their contents. Even 
extremely clean materials can outgas enough to have a measurable effect on 
vacuum quality. Background gases adhere to the surfaces of the chamber or 
diffuse through chamber materials, creating a persistent gas load. 
Heating the chamber or its contents, e.g. a dispenser, 
releases even more gas. We have observed significant gas loads produced
by commercial chromate dispensers when heating them, especially after leaving them at room
temperature for extended periods. For example, in one chamber, a SAES NEXTorr D 
100-5 sputter ion/non-evaporable getter (NEG) hybrid pump normally runs currents 
at the lower limit of observation, between 0 and 1 nA, corresponding to pressure 
at or below $1.5 \times 10^{-11}$ Torr. After 72 hours with the dispenser at 
room temperature, heating the chromate dispenser significantly increases 
the ion pump current. The increase depends on how rapidly the dispenser is 
heated, but currents of 5-6 nA (7.7 - 9.2 $\times 10^{-11}$ Torr) during the 
first hour of heating are common. Longer periods of inactivity seem to
result in larger currents. Whether this is caused by adsorption of waste gases 
onto the steel portion of the dispenser or reversible absorption into the getter 
material is not certain, but it is reasonable to suppose that a dispenser 
without dedicated internal getter material might collect less gas over time. 
The experiment described below tests this hypothesis, comparing 
adsorbed/absorbed gases from IHOPG and chromate dispensers after various periods 
left at room temperature.


In order to measure absorbed gas, the dispensers were left at room temperature 
for a range of times between 2 and 72 hours, and then heated to a constant 
temperature, corresponding to about 1.8 mV of fluorescence, or about $2 \times 
10^{-10}$ Torr of $^{87}$Rb in the glass chamber. During the heating process, 
the RGA measured various gas pressures every few seconds. The resulting curves 
were compared to a curve produced by the dispenser after 1 hour at room 
temperature (i.e., when almost no gas should have been absorbed). The area 
difference between the curves is proportional to the total waste gas released, 
as illustrated in Figure \ref{fig:degas}. The total gas released is approximated 
by multiplying the area between the curves by the pump rate of the ion pump. 
Figure \ref{fig:abs_gas} shows the results of these tests and calculations.

The results show that both dispensers and the sympathetically-heated parts of 
the chamber collect similar amounts of gas up to about 24 hours at room 
temperature. At longer times, the chromate dispenser and the chamber around it 
continue to absorb, but the IHOPG levels off, except for hydrogen. This is 
unsurprising because rubidium intercalated graphite is known to act as a getter 
for hydrogen at room temperature. \cite{Ichimura1992} These results suggest that 
the IHOPG dispensers attract less total waste gas than the chromate dispensers 
when left cold for an extended period, though the amount of the effect which can 
be attributed to the dispenser rather than nearby chamber walls is difficult to 
determine.  However, the data suggest that the IHOPG may have additional utility 
in experiments where long (days or longer) periods of inactivity are expected, 
in addition to their overall lower level of undesired gas output.

\section{Conclusions and Outlook} \label{sec:conclude}

Experimental data and observations show that IHOPG is a suitable source for 
clean rubidium vapor in a cold atom experiment. IHOPGs consist of relatively 
inexpensive and easily available materials, and their production requires 
equipment typically available to atomic physics laboratories. They are operated 
with similar equipment to chromate dispensers, but require less power to operate 
and produce much less waste gas. Handling time in air can be increased to over 
90 minutes with simple post-processing. Under most circumstances, this should be 
enough time to mount the IHOPG and bring the chamber down to rough vacuum. 
Another potential issue with IHOPGs is the low maximum baking temperature, which 
is limited to the activation temperature of the IHOPG, between 125 and 150 
\textcelsius, but these temperatures are often enough to bake out a chamber, 
albeit over a longer period.  Since the expected temperatures needed to 
intercalate lithium and potassium are higher, it is possible that IHOPG 
dispensers using those metals will have higher activation temperatures, which 
would require more power to dispense gas but allow higher maximum bake out 
temperatures.

The increased capacity per mass and volume of IHOPGs is an advantage 
in experiments where long-term operation without service is necessary, such as 
space applications. The high purity of the output vapor reduces load on the 
vacuum pumps, and minimizes the increase in background pressure, which may be 
especially useful in experiments requiring a compact form factor, where, for 
instance, differential pumping schemes might not be feasible.

Since cesium, potassium, and lithium are known to intercalate into HOPG with 
relative ease, they are excellent candidates for dispenser production as well. 
Cesium's applicable thermodynamic characteristics are very similar to that of 
rubidium, so loading and dispensing may work at very similar temperatures. 
Potassium and lithium have a much lower vapor pressure for a given temperature, 
so higher temperatures will almost certainly be required to load and activate 
the dispenser, but these temperatures may still be lower than those required for 
a potassium or lithium chromate dispenser.

The production of IHOPGs is simple and inexpensive. They can be 
produced with equipment readily available to most atomic physics labs. No 
highly technical skills are required to produce them. Therefore, they may find 
applications in undergraduate-level atomic physics experiments, where budgets 
are limited and students have not yet developed extensive technical skill. 
\cite{Wieman1995} As an undergraduate-level experiment, the production of
IHOPGs could be used to instruct students on the use of a glove box,
basic vacuum protocol, and safe handling of alkali metals.

IHOPGs have already been used in-house to load a 2D grating MOT. The atom 
numbers observed in the 2D and 3D grating MOTs were very favorable compared to 
other similar experiments, \cite{Imhof2017, Nshii2013, Vangeleyn2010, 
Esteve2013} despite using less overall laser power than other grating-based 
systems. An IHOPG is currently in use on an experimental apparatus that 
regularly produces Bose-Einstein condensates. The vacuum chamber was changed 
over from one with a chromate dispenser to one with an IHOPG and has experienced 
a noticeable improvement in the lifetime of atoms in the magnetic trap, from 
around 2 seconds to 5 seconds. The lifetime in the IHOPG chamber has remained at 
this level after about 1 year of nearly continuous operation.  This, in addition 
to the data described here, suggest that the integration of IHOPG dispensers 
into systems is practical and can improve their behavior, compared to chromate 
dispensers. Our experimental apparatus currently under construction are moving 
over to IHOPG from chromate dispensers whenever feasible. 

\section{Acknowledgments}

This work was funded by the Air Force Research Laboratory.  We wish to thank Dr.
Greg Pitz and Joshua Key of AFRL for additional help in producing IHOPGs.

\bibliography{preprint}

\end{document}